\documentclass[]{spie}  

\usepackage{amsmath}
\usepackage{amssymb}
\usepackage{rotating}

\usepackage[]{graphicx}

\newcommand{\etal}{{\em et al.\  }}

\newcommand{\bs}[1]{\boldsymbol{#1}}

\newcommand{\halb}{\frac{1}{2}}

\title{Micro-mechanics of multi-phase ferroelectric domain structures} 

\author{Johannes R\"odel 
\skiplinehalf
Technische Universit\"at Dresden, Institute of Materials Science, 01062 Dresden, Germany
}

\authorinfo{Further author information: E-mail: roedel@tmfs.mpgfk.tu-dresden.de\\
\unitlength 1 cm
\begin{picture}(0,0)
\put(0,-0.5){
\parbox[t]{17cm}{
Copyright 2006 Society of Photo-Optical Instrumentation Engineers.
This paper will be/was published in Smart Structures and Materials: Active Materials and Behavior.  SPIE Proceedings vol. 6170, article No. 6,
and is made available as an electronic preprint with permission of SPIE. One print or electronic copy may be made for personal use only. Systematic or multiple reproduction, distribution to multiple locations via electronic or other means, duplication of any material in this paper for a fee or for commercial purposes, or modification of the content of the paper are prohibited.}
}
\end{picture}
}

\begin{document} 
  \unitlength 1 cm

  \maketitle 


\sloppy
\centerline{{\footnotesize manuscript submitted: March 6, 2006}}

\begin{abstract}
High-strain piezoelectric materials are often ceramics with a complicated constitution. 
In particular, PZT is used with compositions near to a so-called morphotropic phase boundary, where not only different variants of the same phase (domains), but different phases may coexist.
Micro-mechanical models for ferroelectric ceramics would be much more realistic, if these effects could be incorporated.

In this paper, we consider the conditions of mechanical and electrical compatibility of ferroelectric domain structures. We are able to address the question of coexistence of different crystallographic phases within the very same crystallite. In general, the spontaneous strain and spontaneous polarization of different phases are not compatible. 
The numerical analysis of the derived relationships are susceptible to the crystallographic description of the phases in question. In this presentation, a simple analysis and analytical, composition dependent fit of strain and polarization of PZT at room temperature for available data are used.

The outlined approach can be used to model the overall behavior of multi-variant and multi-phase crystallites with certain, simplified geometrical arrangements of the constituents.

\end{abstract}



\section{INTRODUCTION}
\label{sect_intro}  

High-strain piezoelectric materials are often materials with a complicated constitution. 
The most interesting materials, PZT ($\rm Pb Zr_x Ti_{1-x} O_3$), $\rm (Pb Mg_{1/3}Nb_{2/3}O_3)_{1-x}-(Pb Ti O_3)_x$  (PMN-PT) and $\rm Pb Zn_{1/3}Nb_{2/3}O_3)_{1-x}-(Pb Ti O_3)_x$ (PZN-PT) have in common, that they are solid solutions which exhibit a so called morphotropic phase boundary.
These materials show extreme material properties, namely high piezoelectric coefficients and are therefore often used at compositions close to the morphotropic phase boundary (MPB).

\begin{figure}
\begin{center}
\unitlength 0.76 cm
\begin{picture}(10,9.12)
\put(1.685,1.35){\includegraphics[width=6.11cm,height=5.45cm]{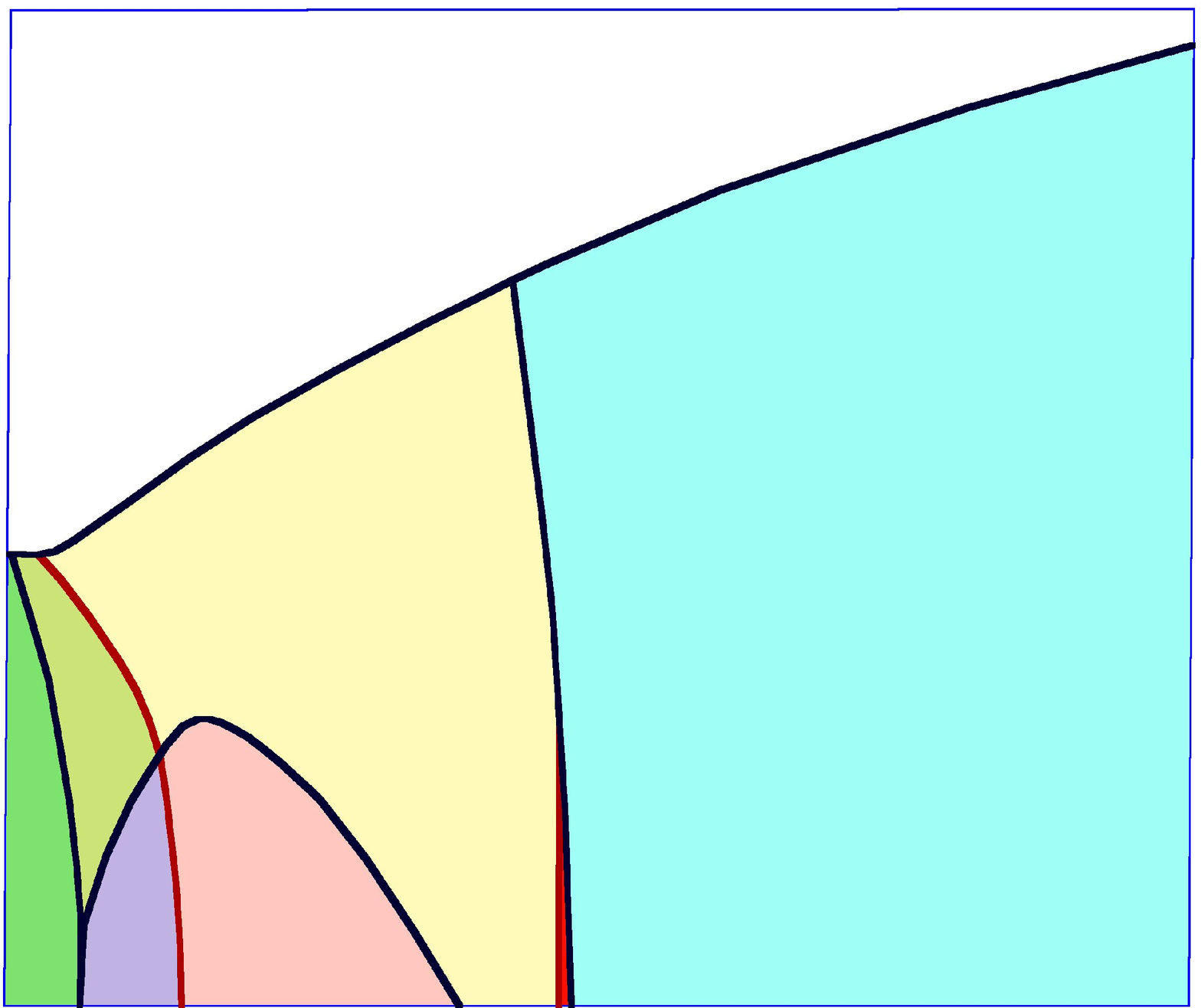}}
\put(0,0){\includegraphics[width=7.8cm]{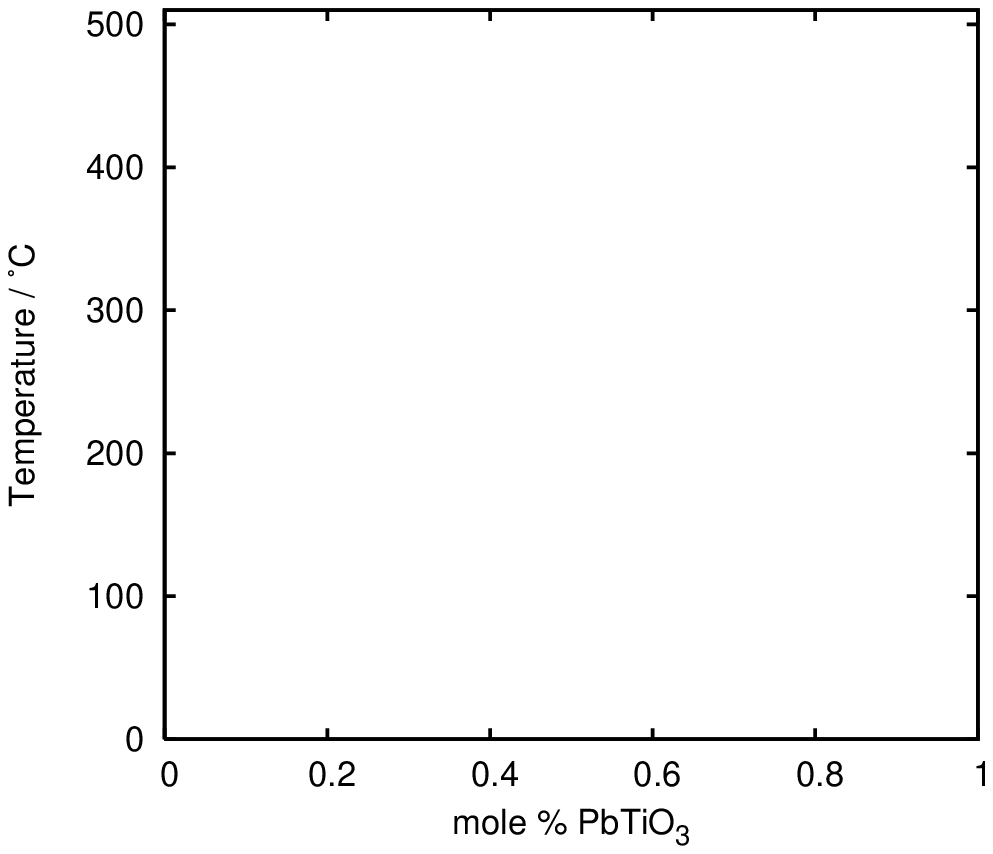}}

\put(4.5,7.5){$\rm \sf Pm\bar 3m$}
\put(7,4.5){\sf P4mm}
\put(6.,2.6){\sf Cm}
\put(3.7,4.5){\sf R3m}
\put(3.2,2.){\sf R3c}
\put(2.15,3.1){\footnotesize \sf Pm}
\put(2.35,1.7){\footnotesize \sf Pc}
\put(1.8,1.6){\begin{turn}{90}\footnotesize \sf Pbam \end{turn}}
\thicklines
\put(6.2,2.4){\vector(-3,-4){0.7}}
\end{picture}
\hspace{1cm}
\begin{picture}(10,10)
\put(0,0){\includegraphics[width=7.6cm]{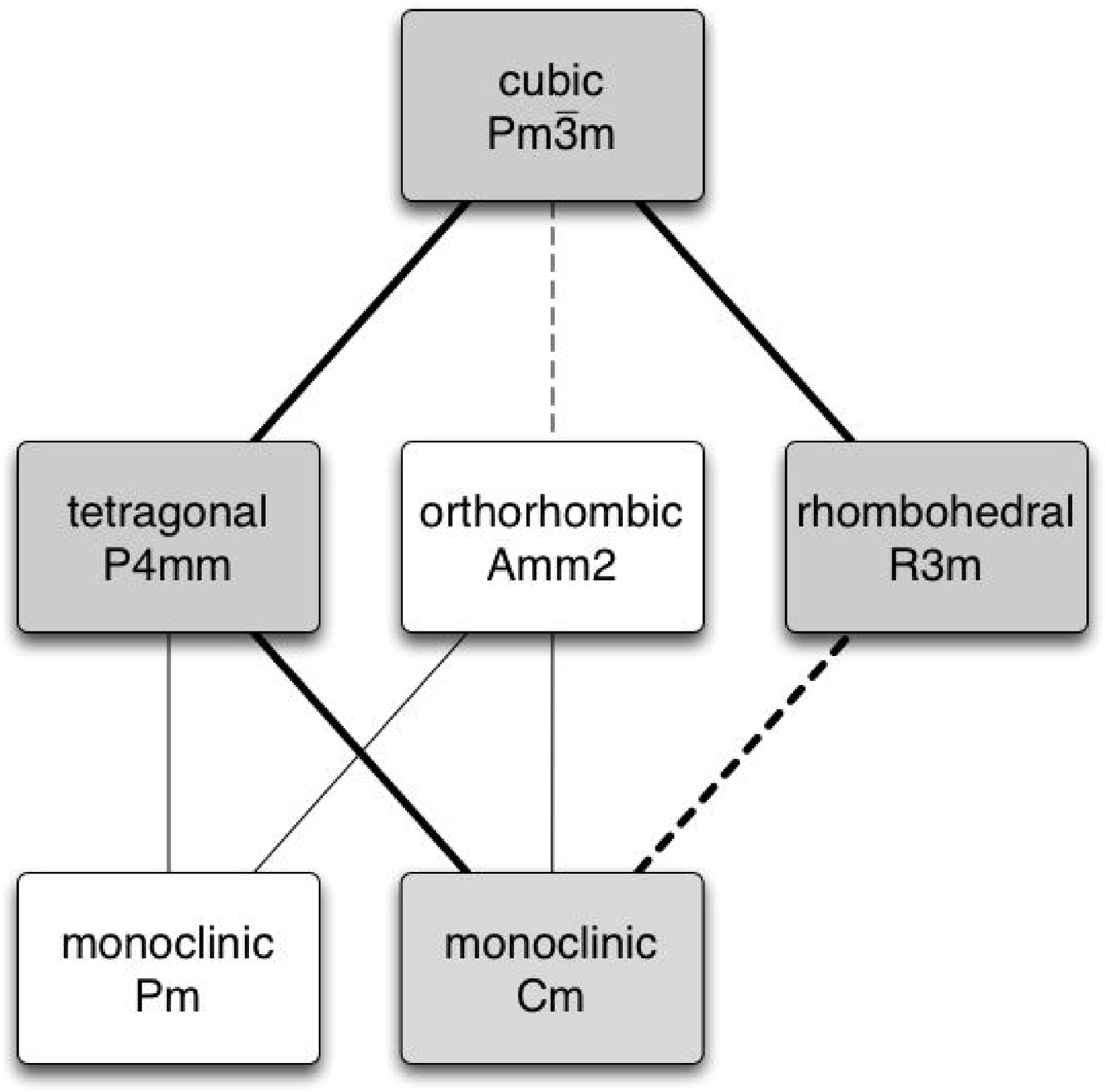}}
\end{picture}

\caption{{\em left: } Schematic phase diagram of PZT, after Ref. \cite{woodward.05}. Besides the traditional cubic, tetragonal and rhombohedral phases, the phase diagram shows schematically also the recently discovered intermediate phases Cm, Pm and Pc. An additional low temperature phase (Cc as an intermediate phase between R3c and Cm) at the morphotropic phase boundary is not shown.
{\em right: } Space group relationships among tilt-free space groups of ABO$_3$ perovskites.  Lines joining two space groups indicate a group-subgroup relationship, where solid lines stand for possible 2nd. order (continuous) transformations and dashed lines indicate first order transformations. After Howard and Stokes, Ref. \cite{howard.05}}
\label{PZT-PD}
\end{center}
\end{figure}

The MPB in the phase diagram (see fig. \ref{PZT-PD}) separates phase regions of two different ferroelectric phases, in the case of PZT the tetragonal and the rhombohedral phase.
The MPB was also considered to be a small region of coexistence of different phases.  
In regard of the exceptional properties of these materials, it is often discussed, that a phase transition of one phase to the other would be possible under applied electric fields, for instance. This would give the material a larger degree of freedom to respond to applied loadings and would explain exceptional electromechanical coupling phenomena.

Less than ten years ago, a closer investigations of PZT at compositions at the MPB \cite{noheda.99,noheda.00,noheda.01} have revealed, that there exists a small region, where a monoclinic phase can be observed. The existence of the monoclinic phase is more pronounced at low temperatures, but it is also observable at room temperature, mostly in coexistence with the tetragonal phase. For a review of the crystal structural studies see Ref.~\citenum{woodward.05}.
It can also be deduced, that the monoclinic phase a intermediate phase between the tetragonal and the rhombohedral phase and that both, tetragonal space group P4mm and rhombohedral space group R3m are subgroups of the monoclinic space group Cm (see fig. \ref{PZT-PD}). This opens the possibility that a tetragonal to rhombohedral phase transition could take place via a monoclinic intermediate state.

More recently, new experiments brought new arguments into the discussion about the nature of the morphotropic phase boundary. Glazer \etal  \cite{glazer.04} have concluded from local electron diffractions, that the local structure of PZT is always monoclinic, and the average tetragonal and rhombohedral structures are a result of short-range to long-range ordered states.
These new viewpoints may stimulate new constitutive models of ferroelectrics with compositions close to a MPB, but so far, a constitutional model, which deals with changes of the structural order is not available. Therefore, micromechanical models will still be a starting point to discuss the behavior of the material. This is also justified by the clearly visible ferroelectric domains which represent ordered states at a mesoscopic level.
For that reasons, the purpose of this paper is to examine the compatibility of tetragonal, rhombohedral and monoclinic domains, depending on the chemical composition of PZT across the MPB.
The starting point of our study is a homogenous crystal which is in a virtual cubic state (state of reference) and may transform to any of the three ferroelectric phases. All three phases may have the same free energy, so that there is no preferred state. A phase transition from one state to the other with a low activation energy is possible if the interface between the two phases are compatible, i.e. if the the deformations and polarization due to the phase transitions (i.e. spontaneous strains and polarizations) are compatible across the interface.
If the spontaneous strains and polarizations are not compatible across the interface, then the compatibility of the kinematic and the dielectric displacements have to be satisfied by elastic strains and 
dielectric displacements, which will give rise to internal fields and associated energy states.

In this work, we will give a brief description of the deformations and polarizations in PZT due to the phase transition. We will then introduce the equations of compatibility, which have to be satisfied and will analyze the three possible transformations: tetragonal$\leftrightarrow$rhombohedral, tetragonal$\leftrightarrow$monoclinic, and monoclinic$\leftrightarrow$rhombo\-hedral.

The micromechanics approach, employed in this work, is that the material may consists of domains with different properties (spontaneous deformations and polarizations), which represent the different phases.
Within the domains, the properties are homogeneous and the material behavior is linear.
The interfaces between domains are sharp interfaces, where the properties may exhibit stepwise changes. The material may show a nonlinear response to applied loads due to moving domain walls or nucleation and growth of new domains.

\section{SPONTANEOUS DEFORMATION AND POLARIZATION}

\begin{figure}[hbtp]
   \begin{center}
   \unitlength 1 cm
   \begin{picture}(10,3.5)\put(0,-1){
   \put(0,0){\includegraphics[width=10cm]{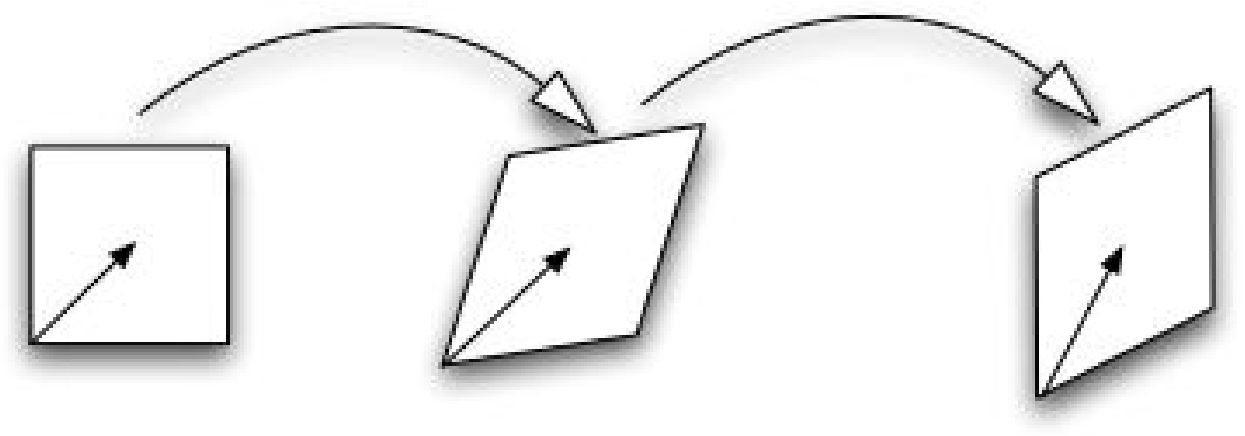}}
   \put(0.8,1.7){\large $\bs x$}
   \put(4.2,1.7){\large $\bs x'$}
   \put(8.2,1.8){\large $\bs x''$}
   \put(2.7,3.5){\large $\bs U$}
   \put(6.5,3.5){\large $\bs Q$}}
   \end{picture}
   \end{center}
   \caption{ \label{fig:def} Schematic of a sequence of deformations. $\bs U$ is the deformation gradient due to the phase transition, $\bs Q$ is a rigid body rotation. The final deformation gradient, $\bs F$, which maps a vector $\bs x$ of the reference configuration onto the vector $\bs x''$ of deformed configuration is $\bs F = \bs Q\bs U$}
   \end{figure}

In the course of a phase transition, the material will be deformed due to changes in the crystal symmety and unit cell geometry.
We will give in brief the deformation gradients due to the phase transitions in question.
The deformation gradient is the linear transformation which maps reference configuration onto a deformed configuration
\[
\bs x' = \bs F \bs x
\]
We distinguish between  the deformation due to the phase transition, $\bs U$, a rigid body rotation, $\bs Q$, and the general deformation gradient, $\bs F$, (see. Fig. \ref{fig:def}). 

\begin{figure}[btp]
   \begin{center}
   \begin{picture}(8,7)
   \put(0,0){\includegraphics[height=7cm]{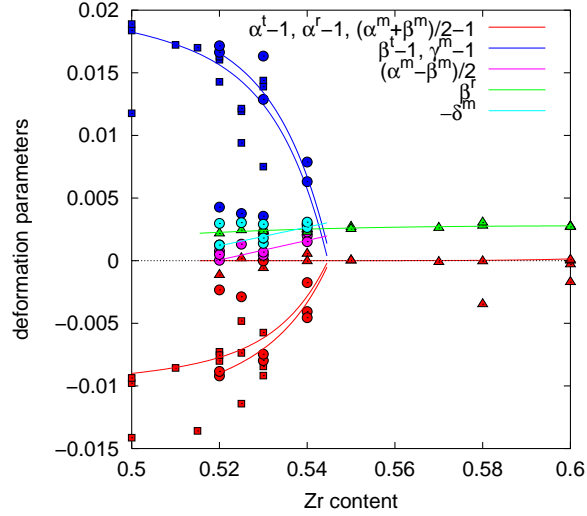}}
   \end{picture}
   \end{center}
   \caption{ \label{fig:PZTdef} Plot of the deformation parameters of the tetragonal, rhombohedral and monoclinic phase of PZT depending on the composition across the MPB}
   \end{figure}

Taking the cubic state as a reference state, the tetragonal, the rhombohedral, and the monoclinic deformation gradients due to phase transitions are
\[
\bs U^{t} = \left(\begin{array}{ccc}
 \alpha^t & 0   & 0  \\
0  & \alpha^t  & 0  \\
0  & 0  & \beta^t  
\end{array}\right),
\qquad
\bs U^r = \left(\begin{array}{ccc}
 \alpha^r & \beta^r  & \beta^r  \\
\beta^r  & \alpha^r  & \beta^r  \\
\beta^r  & \beta^r  & \alpha^r  
\end{array}\right),
\qquad
\bs U^m = \left(\begin{array}{ccc}
 \alpha^m & 0   & \delta^m  \\
0  & \beta^m  & 0  \\
0  & 0  & \gamma^m
\end{array}\right)
\]
It should be noted, that these deformation gradients are given in different coordinate systems.
While the tetragonal and rhombohedral deformation is defined in a coordinate system, which is parallel to principle axis of the cubic cell, the monoclinic deformation is given in a coordinate system with principle axis $x_1$ is parallel to the pseudo-cubic [110] direction, and $x_3 || [001]^{\rm pc}$.
To use the deformation gradients on a common basis, either the rhombohedral, or the monoclinic deformation gradients have to be rotated by 45$^\circ$ about the $[110]^{pc}$ axis.
As a result, the monoclinic deformation gradient in pseudo-cubic(ally oriented) coordinates is
\[
\bs U^m = \left(\begin{array}{ccc}
\halb(\alpha^m+\beta^m) & \halb(\alpha^m-\beta^m)    & \frac{1}{\sqrt{2}}\delta^m  \\
\halb(\alpha^m-\beta^m)   &\halb(\alpha^m+\beta^m)  & \frac{1}{\sqrt{2}}\delta^m   \\
0  & 0  & \gamma^m
\end{array}\right),
\] 
or the rhombohedral deformation gradient in the rotated coordinate system is
\[
\bs U^r = \left(\begin{array}{ccc}
 \alpha^r+ \beta^r  & 0 & \sqrt{2}\beta^r  \\
0 & \alpha^r -\beta^r  & 0  \\
\sqrt{2}\beta^r  & 0  & \alpha^r  
\end{array}\right)
\]
The polarizations of the tetragonal, rhombohedral and monoclinic state in the global coordinate system parallel to the pseudo-cubic axis are
\[
\bs P^t = \left(\begin{array}{c}
0 \\ 0 \\ P^t
\end{array}\right), 
\qquad
\bs P^r = \left(\begin{array}{c}
P^r \\ P^r \\ P^r
\end{array}\right), 
\qquad
\bs P^r = \left(\begin{array}{c}
P_1^m \\ P_1^m \\ P_3^m
\end{array}\right)
\]
The parameters of the deformation gradients have been fitted to data according available structural investigations. The polarizations can be estimated from deformations and sublattice shifts, but will not be given here. The results for the deformations have been plotted in Fig. \ref{fig:PZTdef} and details will be published elsewhere\cite{roedel-PZT-paper}.

\section{Compatibility of domains}

\begin{figure}[btp]
   \begin{center}
   \begin{picture}(10,4)
  \put(0,0){ \includegraphics[width=10cm]{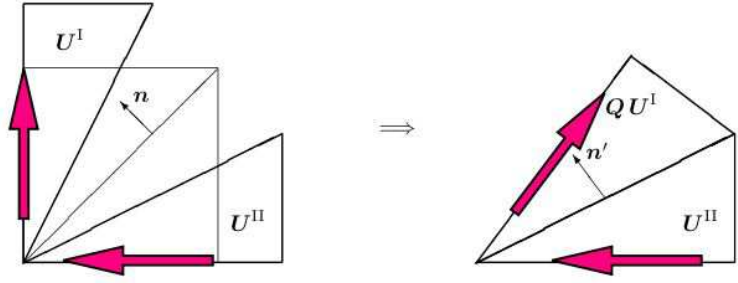}}
   \end{picture}
   \end{center}
   \caption{ \label{fig:def-comptbl} Schematic of a crystal divided into two domains with different deformation gradients. To make the two domains compatible along an interface with normal $\bs n$ (in reference configuration), or $\bs n'$ (in the deformed configuration) at least one of two domains must be rotated by the rotation $\bs Q$. }
   \end{figure}

Consider a crystal which volume can be divided into two domains with different deformation gradients (see Fig. \ref{fig:def-comptbl}).
The crystal is considered to be under traction- and field-free boundary condition, so that there are no internal fields (stresses and electric fields) in the pre-transformed state, and also no internal fields if the transformation is compatible.
Given the deformation gradients of the two domains, we want to know, if these two deformations can be compatible. If so, what is the normal direction of the interface? In general, there are necessary rigid body rotations, $\bs Q$, to make the two domains compatible (see Fig. \ref{fig:def-comptbl}).

The deformation gradients of the two domains I and II have to satisfy the following equation of kinematic compatibility\cite{bhattacharya.martensit}:
\[
\bs F^I - \bs F^{II} = \bs b \otimes \bs n
\]
For a detailed discussion of this equation see for instance Ref. \citenum{bhattacharya.martensit}.
If we allow domain I to be rotated by the rotation $\bs Q$ and insert the deformations due to the phase transitions $\bs U^I$ and $\bs U^{II}$ the equation of kinematic compatibility becomes
\[
\bs Q \bs U^I - \bs U^{II} = \bs b \otimes \bs n
\] 
$\bs b$ has to be a vector with real components and $\bs n$ is the normal of the interface between the two domains in the reference state.
If the material is a ferroelectric material, then also the polarizations have to be compatible. This means, that the normal component of the polarizations should be continuous across the interface $\bs n'$, i.e. it should be compatible after the transformation.
Taking into account, that the polarizations are correlated to the orientation of the crystal lattice, and due to the deformations the vector $\bs n$ is no longer a normal to the interface, the equation of compatibility of the polarization reads\cite{shu.01}:
\[
(\kappa \,\bs Q \bs P^{s\,I} - \bs P^{s\,II}) \,{\bs U^{\rm II}}^{-T} \bs n = 0 
\]
where $\kappa = \pm 1$, taking into account, that there are usually two opposite polarizations states associated with one deformation state.

Ball and James \cite{ball.87} have found an algorithm, which provides a general solution for the problem of kinemetic compatibility \cite{bhattacharya.martensit}, which will be briefly summarized: 
\begin{enumerate}
\item  Define matrix $\bs C = {\bs U^{\rm II}}^{-t}{\bs U^{\rm I}}^{t} 
\bs U^{\rm I}{\bs U^{\rm II}}^{-1} $

\item If $\bs C = \bs I$ $\to$ the problem has no solution

\item If $\bs C \neq \bs I$, calculate the eigenvalues 
$\lambda_1,\; \lambda_2,\;\lambda_3$ of the  matrix $\bs C$. \\
$\lambda_i >0$, $\to$ $\lambda_1\leq \lambda_2\leq\lambda_3$.

\item A solution exists if 
$
\lambda_1 \leq 1, \quad \lambda_2 = 1, \quad \lambda_3 \geq 1
$
\item Then, there are two solutions:
\begin{eqnarray*}
\bs b &=& \rho \left(
\sqrt{\frac{\lambda_3 (1-\lambda_1)}{\lambda_3-\lambda_1} \hat{\bs e}_1}
+ \kappa \sqrt{\frac{\lambda_1 (\lambda_3-1)}{\lambda_3-\lambda_1} \hat{\bs e}_3}
\right)\\
\bs n &=& \frac{\sqrt{\lambda_3}-\sqrt{\lambda_1}}
{\rho \sqrt{\lambda_3-\lambda_1}}\left(-\sqrt{1-\lambda_1}{\bs U^{\rm II}}^t \hat{\bs e}_1
+ \kappa \sqrt{\lambda_3-1}{\bs U^{\rm II}}^t \hat{\bs e}_3 
\right)\\
\kappa &=& \pm 1
\end{eqnarray*} 
$\rho$ is choosen so that $|\bs n| =1$ and $\hat{ \bs e}_i$ are the normalized eigenvectors, corresponding to eigenvalues $\lambda_i$
\item 
The necessary rotation of domain I is obtained by inserting $\bs b$ and $\bs n$ into the compatibility equation:
$\displaystyle
\bs Q = \left(\bs b \otimes \bs n + \bs U^{\rm II} \right)\; {\bs U^{\rm I}}^{-1}
$
\end{enumerate}

This scheme provides the complete solution for the kinematic compatibility of two domains with arbitrary deformations.
Therefore, the electromechanical problem which requires both, kinematic and electrical compatibility is over-determined and has in general no solution.
It can be shown, that a solution for the coupled problem exits only, if the the two domains are symmetry related \cite{desimone.02}. I.e., only if the two domains are different variants of the same crystallographic phase, then both the deformations and the polarization can be compatible. This is what would be expected and which gives the well known relationships for the non-180$^\circ$ domains in various ferroelectrics.

On the other hand, this also means that the transition from one ferroelectric phase to another can not be fully compatible, no matter which particular transition is considered.
For that reason, here we consider only the kinematic compatibility. If a compatible interface plane is found, the polarization charges, which appear on an electrical incompatible interface, can be calculated (see Sect. \ref{outlook}).

\section{RESULTS}

In order to analyze the possibility of a compatible phase transition, the eigenvalues of the matrices  $\bs C$ according to the various possible transitions have to be analyzed.
A compatible transition is possible, if one of the eigenvalues is equal to 1.
For the tetragonal$\leftrightarrow$rhombohedral transition it is found, that this condition is fulfilled, if
\[
\alpha^t = \alpha^r-\beta^r.
\]
The tetragonal$\leftrightarrow$monoclinic transition is compatible, if 
\[
\alpha^t = \beta^m,
\]
and the rhombohedral$\leftrightarrow$monoclinic transition is compatible if
\[
\beta^m = \alpha^r-\beta^r.
\]
The deformation parameters according to these three conditions have been plotted versus the composition across the MPB in Figs. \ref{fig:komptbl-tr} -  \ref{fig:komptbl-rm}. 
It is found, that the tetragonal$\leftrightarrow$rhombohedral may be compatible at a composition at the very Zr-rich side of the MPB ($x \approx 0.54$). However, in this compositional region, there are no experimental data for the tetragonal phase, so that it may not exist at this Zr-concentrations.
On the other hand, the tetragonal$\leftrightarrow$monoclinic transition is never compatible according to the accessible data and the monoclinic$\leftrightarrow$rhombohedral transition may be compatible at even higher Zr-contents as  tetragonal$\leftrightarrow$rhombohedral.

In general, it can be concluded that one can find an interface between the two domains with comaptible
deformation, if the out-of-plane deformations in the $\rm [\bar 1 10]^{pc}$ plane are homogeneous.
Then, there is a compatible interface with a normal vector in the $\rm [\bar 1 10]^{pc}$ plane.

Taking for instance the differences $\alpha^t = \beta^m$ as a measure of incompatibility for the transition, it can be noted that the incompatibility between the tetragonal and the monoclinic phase is almost constant across the MPB, but relatively low compared to the other transitions. This may explain, why tetragonal and monoclinic phases seem to coexist within the MPB region.

\unitlength 1 cm

   \begin{figure}[b]
   \begin{center}
   \begin{picture}(10,7)
   \includegraphics[height=7cm]{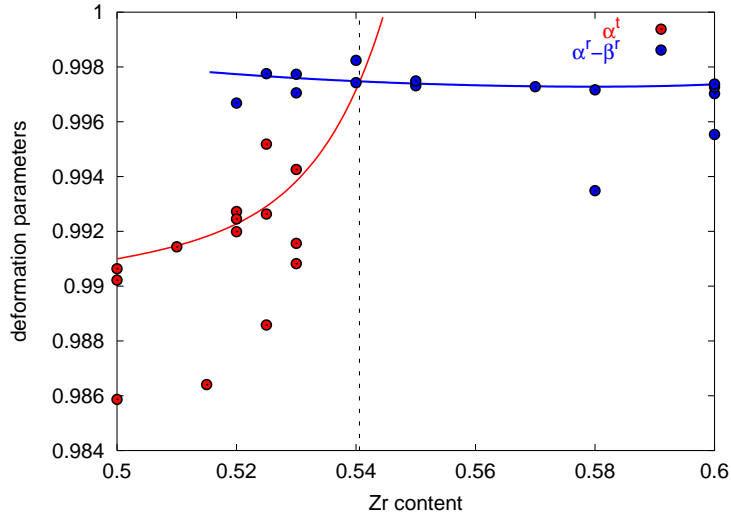}
   
   \end{picture}
   \end{center}
   \caption[example] 
     { \label{fig:komptbl-tr} Plot of the deformation parameters $\alpha^t$ and $\alpha^r-\beta^r$ versus the composition of PZT. At the intersection of both curves, the tetragonal and the rhombohedral phase could be compatible}
   \end{figure}

\section{outlook}\label{outlook}
  
The micromechanical approach to phase transitions in ferroelectrics at the MPB can be extended towards energy minimizations for incompatible interfaces. 
The internal fields arising due to incompatible deformations and polarizations can be modeled by distributed interface defects.
For polarizations it is  obvious, that the charge density at the interfaces can be computed from the differences of the polarizations:
\[
\rho = ( \,\bs Q \bs P^{s\,I} - \bs P^{s\,II}) \,{\bs U^{\rm II}}^{-T} \bs n 
\]
The analogous model for the kinematic defects are distributed dislocations, expressed by the interface dislocation density tensor \cite{mura}
\[
\alpha_{ij} =  - e_{ikm} n_k (F_{mj}^{\rm II}-F_{mj}^{\rm I})
\]

In order to calculate the energy contributions of both types of defects and to search a plane with minimal energy, it is necessary to know the complete geometry of the domains as well as the the material properties. This is far beyond the scope this contribution.

   \begin{figure}
   \begin{center}
   \begin{picture}(10,7)
   \includegraphics[height=7cm]{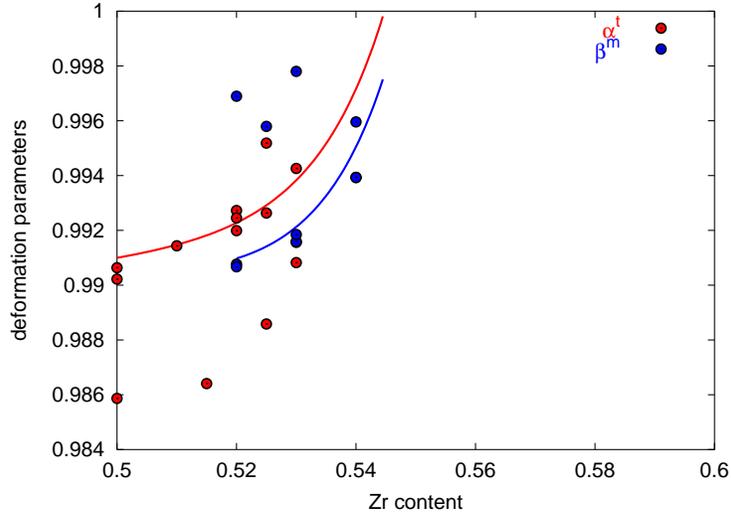}
   \end{picture}
   \end{center}
   \caption[example] 
     { \label{fig:komptbl-tm} Plot of the deformation parameters $\alpha^t$ and $\beta^m$ versus the composition of PZT. As the two curves do not intersect, the tetragonal and the monoclinic phase are not compatible}
   \end{figure}

   \begin{figure}
   \begin{center}
   \begin{picture}(10,7)
   \includegraphics[height=7cm]{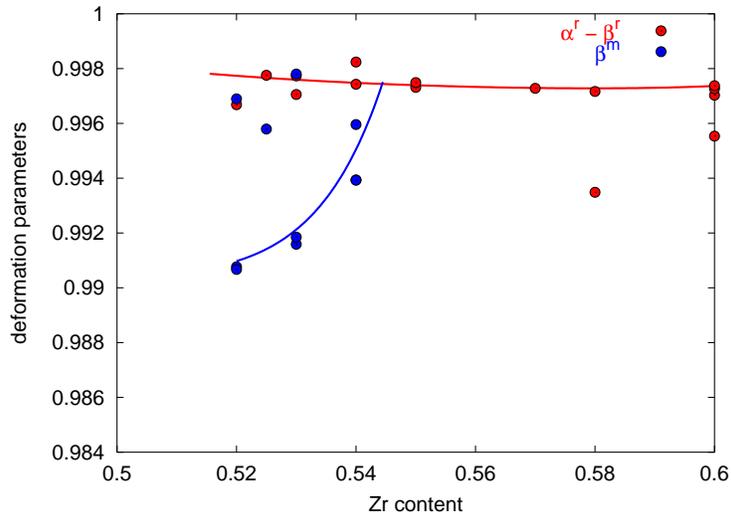}
   \end{picture}
   \end{center}
   \caption[example] 
     { \label{fig:komptbl-rm} Plot of the deformation parameters $\beta^m$ and $\alpha^r-\beta^r$ versus the composition of PZT. At the intersection of both curves, the monoclinic and the rhombohedral phase could be compatible}
   \end{figure}

\section{summary}\label{summary}

The present study presents a micromechanical approach to phase transitions in ferroelectric materials at a morphotropic phase boundary. The peculiar property of such kind of materials is that there is the possibility of transitions from one ferroelectric phase to another. This transitions have low energy barriers and can then increase the degree of freedom to respond on applied loads, if the interfaces  between different phases are compatible.
The compatibility problem has been studied using a well established formalism. It should be noted, that there exist in general no solution for the entire electromechanical problem, where both deformations and polarization have to be compatible. For that reason, only the kinematic compatibility is considered here.
For the particular deformation parameters of PZT it is found that the tetragonal$\leftrightarrow$monoclinic transition is very close, but never exactly compatible. The 
tetragonal$\leftrightarrow$rhombohedral transition has a point of compatibility at $x \approx 0.54$ and the rhombohedral$\leftrightarrow$monoclinic transition has a point of compatibility at even higher Zr contents.
This study helps to explain why the tetragonal and the monoclinic phase may coexist in the entire range of the MPB. But all phase transitions in question are associated with incompatible phase interfaces which requires to overcome a certain critical energy.

The outlined approach can be used to model the overall behavior of multi-variant and multi-phase crystallites with certain, simplified geometrical arrangements of the constituents.

\acknowledgments     
 
This financial support of this work by the Deutsche Forschungsgemeinschaft (DFG) is gratefully acknowledged.


\bibliography{/users/roedel/tex/bib/piezo,/users/roedel/tex/bib/microstr,/users/roedel/tex/bib/elast}   

\begin{thebibliography}{10}

\bibitem{woodward.05}
D.~I. Woodward, J.~Knudsen, and I.~M. Reaney, ``Review of crystal and domain
  structures in the {$\rm PbZr_xTi_{1?x}O_3$} solid solution,'' {\em Phys. Rev.
  B}~{\bf 72}, p.~104110, 2005.

\bibitem{howard.05}
C.~J. Howard and H.~T. Stokes, ``Structures and phase transitions in
  perovskites - group-theoretical approach,'' {\em Acta Cryst.}~{\bf A 61},
  pp.~93--11, 2005.

\bibitem{noheda.99}
B.~Noheda, D.~E. Cox, G.~Shirane, J.~A. Gonzalo, L.~E. Cross, and S.-E. Park,
  ``A monoclinic ferroelectric phase in the {PbZr$_{1-x}$Ti$_{x}$O$_3$} solid
  solution,'' {\em Appl. Phys. Letters}~{\bf 74}(14), pp.~2059--2061, 1999.

\bibitem{noheda.00}
B.~Noheda, J.~A. Gonzalo, L.~E.Cross, R.~Guo, S.-E. Park, D.~E. Cox, and
  G.~Shirane, ``Tetragonal-to-monoclinic phase transition in a ferroelectric
  perovskite: the structure of {PbZr$_{0.52}$Ti$_{0.48}$O$_3$},'' {\em Phys.
  Rev. B}~{\bf 61}, pp.~8687--8695, 2000.

\bibitem{noheda.01}
B.~Noheda, D.~E. Cox, G.~Shirane, R.~Guo, L.~E.Cross, B.~Jones, and D.~E. Cox,
  ``Stability of the monoclinic phase in ferroelectric perovskite
  {PbZr$_{1-x}$Ti$_{x}$O$_3$},'' {\em Phys. Rev. B}~{\bf 63}, p.~014103, 2001.

\bibitem{glazer.04}
A.~M. Glazer, P.~A. Thomas, K.~Z. Baba-Kishi, G.~K.~H. Pang, and C.~W. Tai,
  ``Influence of short-range and long-range order on the evolution of the
  morphotropic phase boundary in {$\rm Pb Zr_{1-x}Ti_x O_3$},'' {\em Phys. Rev.
  B}~{\bf 70}, p.~184123, 2004.

\bibitem{roedel-PZT-paper}
J.~R\"odel, ``Spontaneous strain and polarization of {$\rm
  Pb(Zr_xTi_{1-x})O_3$} -- a topical review of crystal structures at room
  temperature.'' in preparation, 2006.

\bibitem{bhattacharya.martensit}
K.~Bhattacharya, {\em Microstructure of Martensite}, Oxford University Press,
  New York, 2003.

\bibitem{shu.01}
Y.~C. Shu and K.~Bhattacharya, ``Domain patterns and macroscopic behaviour of
  ferroelectric materials,'' {\em Phil. Mag. B}~{\bf 81}, pp.~2021--2054, 2001.

\bibitem{ball.87}
J.~M. Ball and R.~D. James, ``Fine phase mixtures as minimizers of energy,''
  {\em Arch. Rat. Mech. Anal.}~{\bf 100}, pp.~13--52, 1987.

\bibitem{desimone.02}
A.~DeSimone and R.~D. James, ``A constrained theory of magnetoelasticity,''
  {\em J. Mech. Phys. Solids}~{\bf 50}, pp.~283--320, 2002.

\bibitem{mura}
T.~Mura, {\em Micromechanics of Defects in Solids}, Kluwer, Dordrecht, 2.~ed.,
  1993.

\end{thebibliography}
\bibliographystyle{spiebib}   

\end{document}